\documentclass[preprint, nofootinbib]{revtex4}

\usepackage[]{caption}
\usepackage{color, graphicx, bm}
\usepackage{dcolumn}% Align table columns on decimal point
\usepackage{subfig}
\usepackage{enumerate}
\usepackage{simplewick}  % Wick contraction

\def\be{\begin{equation}}
\def\ee{\end{equation}}
\def\bea{\begin{eqnarray}}
\def\eea{\end{eqnarray}}
\def\dprime{\prime\prime}
\def\calh{\mathcal{H}}

\def\cals{\mathcal{S}}

\def\super3{{}^{(3)}}

\def\calr{\mathcal{R}}
\def\calc{\mathcal{C}}

\def\cala{\mathcal{A}}

\def\call{\mathcal{L}}

\def\cald{\mathcal{D}}

\def\calv{\mathcal{V}}

\begin{document}

\title{Hamiltonian analysis of Linearized Extension of
Ho\v{r}ava-Lifshitz gravity}

\author {Seoktae Koh\footnote{E-mail address:
steinkoh@sogang.ac.kr} and Sunyoung Shin\footnote{E-mail address:
shinsy@sogang.ac.kr}}

\affiliation{Center for Quantum Spacetime, Sogang University, Seoul
121-742, Republic of Korea}

%\date{\today}
\begin{abstract}
We investigate the Hamiltonian structure of linearized extended Ho\v{r}ava-
Lifshitz gravity  in a flat cosmological background
following the Faddeev-Jackiw's
Hamiltonian reduction formalism. The Hamiltonian structure
of extended Ho\v{r}ava-Lifshitz gravity is similar to
that of the projectable version of original Ho\v{r}ava-Lifshitz gravity,
in which there is one primary constraint and so
there are two physical degrees of freedom.
We also find that  extra scalar graviton mode in an
inflationary background can
be decoupled from the matter field in the infrared (IR) limit,
but it is coupled to the matter field in
a general cosmological background. But it is necessary
to  go beyond linear order in order to draw any conclusion
of the strong coupling problem.
\end{abstract}
%\pacs{}
% \keywords{}

\maketitle
%%%%%%%%%%%%%%%%%%%%%%%%%%%%%%%%%%%%%%%%%%%%%%%%%%%%%%%%%%%%%%%%%%%%%%%%%%%%%%
%%%%%%%%%%%%%%%%%%%%%%%%%%%%%%%%%%%%%%%%%%%%%%%%%%%%%%%%%%%%%%%%%%%%%%%%%%%%%%
%\section{Introduction}

The reconciliation of gravity and quantum theory, which is important to
understand the very early stage of our Universe and the black hole,
is a very challenging task in theoretical physics.
Among the several proposals of quantum theory of gravity, recently
Ho\v{r}ava\cite{Horava:2009uw} proposed a UV complete, non-relativistic
gravity theory which is
power-counting renormalizable giving up the Lorentz invariance.
Since then,
 many paid attention to this scenario to
apply to the black hole\cite{Lu:2009em}, cosmology
\cite{Calcagni:2009ar}\cite{Mukohyama:2010xz}
and observational tests\cite{Park:2009zr}.
In spite of its many appealing properties, it seems to suffer from
many problems \cite{Blas:2009yd}\cite{Li:2009bg}
 such as instability, strong coupling, renormalizability etc.

In order to alleviate the original Ho\v{r}ava-Lifshitz gravity's
problem, in \cite{Blas:2009qj} the extended version of
Ho\v{r}ava-Lifshitz gravity is proposed, in which the new degree of
freedom by taking the spatial gradient of a lapse function is
introduced without violating the symmetry of the action. They argued
that the strong coupling problem of the scalar graviton mode in the
IR limit would be solved. There still remains some debate on strong
coupling problem in extended version as well as on the physical degrees of
freedom \cite{Kluson:2010nf}\cite{Blas:2010hb}.

In this paper, following the previous work\cite{Gong:2010xp}
of one of the authors,
we investigate the Hamiltonian structure of  linearized extended
Ho\v{r}ava-Lifshitz gravity in a cosmological background using the
Faddeev-Jackiw approach \cite{Faddeev:1988qp}. First, we derive the quadratic
 action in the extended Ho\v{r}ava-Lifshitz gravity model and then obtain
constraints and Hamiltonian. Through the investigation of the Poisson algebra,
the physical degrees of freedom are exactly counted
 and we analyze the Hamiltonian structures.
Next, we obtain the equations of motion of the physical degrees of freedom
and finally we briefly comment about the strong coupling issues in our
case in the IR limit.

%%%%%%%%%%%%%%%%%%%%%%%%%%%%%%%%%%%%%%%%%%%%%%%%%%%%%%%%%%%%%%%%%%%%
%\section{Extended Ho\v{r}ava-Lifshitz gravity}
%%%%%%%%%%%%%%%%%%%%%%%%%%%%%%%%%%%%%%%%%%%%%%%%%%%%%%%%%%%%%%%%%%%%
%%%%%%%%%%%%%%%%%%%%%%%%%%%%%%%%%%%%%%%%%%%%%%%%%%%%%%%%%%%%%%%%%%%%
%\subsection{Gravity action}
%%%%%%%%%%%%%%%%%%%%%%%%%%%%%%%%%%%%%%%%%%%%%%%%%%%%%%%%%%%%%%%%%%%%

We consider the Arnowitt-Deser-Misner (ADM) metric which is given by
\bea%
ds^2 = (-N^2 + N_i N^i)dt^2 +2N_i dt dx^i +\gamma_{ij}dx^i dx^j,
\label{eq:adm}
\eea%
where $N$ is the lapse function, $N_i$ are shift
vectors, and $\gamma_{ij}$ is the spatial 3 metric. The %extended
Ho\v{r}ava-Lifshitz gravity action  in
 the ADM metric with a single
scalar field is
\bea%
\cals = \int d^4x N\sqrt{\gamma} \biggl[\frac{1}{2\kappa^2}(
K_{ij}K^{ij} -\lambda K^2) - \calv + \frac{1}{2N^2}\biggl(
\dot{\phi} - N^i \partial_i \phi \biggr)^2 - Z (\phi) -V(\phi)
\biggr] \label{eq:grav_act}.
\eea%
The extrinsic
curvature $K_{ij}$ and its trace are written in terms of the ADM
metric (\ref{eq:adm}) as
\bea%
K_{ij} &=& \frac{1}{2N} \left(\partial_i N_j + \partial_j N_i -
\frac{\partial \gamma_{ij}}{\partial t} \right),
\label{eq:extrinsic} \\
K &=& \gamma^{ij}K_{ij} = K^i_i. \label{eq:textrinsci}
\eea%
For extended Ho\v{r}ava-Lifshitz
 gravity \cite{Blas:2009qj}, the gravitational potential term $\calv$  in the action (\ref{eq:grav_act}) depends on $\gamma_{ij}$, its spatial derivative
 and on 3-dimensional vector $a_i$ constructed from the
 lapse function $N(t,{\bf x})$ as
\bea%
a_i = \frac{\partial_i N(t,{\bf x})}{N(t,{\bf
x})},
\eea%
which represents the proper acceleration of the vector field of unit
normals to the foliation surfaces \cite{Blas:2009yd}.
Under the anisotropic scaling transformations
\bea
{\bf x} \rightarrow l {\bf x}, \quad t \rightarrow l^{z} t,
\eea
the $z =3$ theory in the UV is power-counting renormalizable,
so the potential in the action (\ref{eq:grav_act}) can have
at most 6-th order spatial derivative terms.
With these spirits, the gravitational potential for extended
Ho\v{r}ava-Lifshitz gravity  can take the form
\bea%
\calv &=& -\xi R - \alpha a_i a^i + f_1 R^2 + f_2 R_{ij} R^{ij} +
f_3 R \partial_i a^i + f_4 a_i \partial^2 a^i \nonumber \\
& & + g_1 (\partial_i R)^2 + g_2 \partial_i R_{jk} \partial^i R^{jk}  +
g_3 \partial^2 R \partial_i a^i + g_4 a_i \partial^4 a^i, \label{eq:ehl_pot}
\eea%
where $\xi, \alpha, f_{n}, g_{n}$ are constants
 and $\partial^2 = \partial_i \partial^i$.
 $R$ and $R_{ij}$ are 3-dimensional Ricci scalar and Ricci tensor, respectively.
Since we are interested in the linear analysis of
extend Ho\v{r}ava-Lifshitz gravity, the only terms in
 (\ref{eq:ehl_pot}) relevant to
the linear analysis on a flat cosmological background are included
\cite{Blas:2009qj}\cite{Kobayashi:2010eh}.
Most general gravitational potential form in
extended Ho\v{r}ava-Lifshitz gravity can be found in Ref. \cite{Blas:2009qj}.
It is known that the action (\ref{eq:grav_act}) with
the gravitational potential (\ref{eq:ehl_pot}) is invariant under the
foliation conserving transformations
\bea
{\bf x} \rightarrow {\bf x} + g(t,{\bf x}), \quad t \rightarrow
t + f(t).
\label{eq:folitransf}
\eea
The $Z(\phi)$ in (\ref{eq:grav_act}) is the matter part potential constructed from the spatial derivative
of a scalar field, which is given by
\bea %
Z(\phi) = \sum_{n=1}^3 \xi_n \partial_i^{(n)} \phi
\partial^{i(n)} \phi.
\eea%
The superscript $(n)$ denotes the $n$-th spatial derivative.
%In
%$z=3$ Ho\v{r}va-Lifshitz gravity, $Z$ keep at most cubic spatial
%derivatives.

In order to derive the background and linear perturbation equations of motion
by varying the action, we expand the metric and the scalar field to the linear
order as
\bea%
N &=& a(\eta)(1+ \Phi),\nonumber\\
\quad N_i &=& a(\eta)^2 \partial_i \beta,
\nonumber \\
\gamma_{ij} &=& a(\eta)^2 (\delta_{ij} + h_{ij}) = a(\eta)^2 \biggl(
(1-2\calr) \delta_{ij} + 2 \biggl(\partial_i \partial_j -
\frac{1}{3}\delta_{ij} \partial^2\biggr) E \biggr),
\label{eq:lmetric2}
\eea%
and
\bea %
\phi(\eta,x)=\phi_0(\eta)+\delta\phi(\eta,x),
\eea%
where the parameter $a(\eta)$ is a scale factor and $\eta$
is a conformal time. In this paper, we only consider
the scalar mode perturbations.

From the linear order of the action (\ref{eq:grav_act})
\bea %
\delta_1 \cals 
&=&\int d^4x a^2 \biggl[ \biggl\{ -\frac{3(1-3\lambda)}{2\kappa^2}
\calh^2 -\frac{1}{2} \phi_0^{\prime 2} - a^2 V_0 \biggr\} \Phi +
\biggl\{ -\frac{(1-3\lambda)}{4\kappa^2} (\calh^2 + 2\calh^{\prime})
\nonumber \\
& & + \frac{1}{2} \biggl(\frac{1}{2}\phi_0^{\prime 2} - a^2 V_0
\biggr) \biggr\} h^k_k + \biggl\{ - \phi_0^{\dprime} -2
\calh\phi_0^{\prime} - a^2 V_{\phi} \biggr\} \delta \phi  \biggr],
\eea%
where $\calh = \frac{a^{\prime}}{a}$ and the prime denotes the
derivative with respect to $\eta$,
we can obtain the  background equation of motion in a flat
cosmological background
\bea %
& & \frac{3(1-3\lambda)}{2\kappa^2}
\calh^2 = -(\frac{1}{2}\phi_0^{\prime 2} + a^2 V_0),
\\
& & \frac{(1-3\lambda)}{2\kappa^2}(\calh^2 + 2\calh^{\prime}) =
\frac{1}{2}\phi_0^{\prime 2} - a^2 V_0,
\\
& & \phi_0^{\dprime} + 2\calh \phi_0^{\prime} + a^2 V_{\phi} = 0.
\eea%

By expanding the action up to 2nd order in terms of the perturbed quantities,
the quadratic action yields
\bea%
\delta_2 \cals
&=&\int d^4x a^2 \biggl[ \frac{1}{2\kappa^2}(1-3\lambda) \biggl\{
3\calh^2 \Phi^2 + 2\calh \Phi \partial^2 (\beta -E^{\prime}) + 6
\calh \Phi\psi^{\prime} + 3\psi^{\prime 2} + 2\psi^{\prime}
\partial^2 (\beta - E^{\prime}) \biggr\}
\nonumber \\
& & +\frac{1}{2\kappa^2}(1-\lambda)  [\partial^2 (\beta
-E^{\prime})]^2 - \alpha \Phi \partial^2 \Phi
 - \frac{1}{a^2}(\partial^2 \psi)^2 (16f_1 + 6f_2)
 - f_3 \frac{4}{a^2}\partial^2\psi\partial^2\Phi
\nonumber \\
& & + \frac{f_4}{a^2} \partial^2 \Phi\partial^2 \Phi
+\frac{1}{a^4}(16 g_1 + 6g_2)\partial^2\psi(\partial^2)^2\psi  - g_3
\frac{4}{a^4} (\partial^2)^2\psi\partial^2\Phi + g_4 \frac{1}{a^4}
\partial^2 \Phi \partial^2 \partial^2 \Phi
\nonumber \\
& &  +  2\xi(2\Phi -\psi) \partial^2 \psi +\frac{1}{2} \delta
\phi^{\prime 2} - \phi_0^{\prime} \Phi \delta \phi^{\prime}
 + \frac{1}{2}\phi_0^{\prime 2} \Phi^2  - a^2 \delta Z
- \frac{1}{2}a^2 V_{\phi\phi} (\delta \phi)^2 -a^2 V_{\phi} \Phi
\delta \phi
\nonumber \\
& &  + 3 \phi_0^{\prime} \psi^{\prime}\delta \phi
+\phi_0^{\prime}\delta \phi \partial^2 (\beta -E^{\prime})
\biggr],\label{eq:totalquadaction}
\eea%
where $\psi$ is defined as \bea \psi = \calr + \frac{1}{3}\partial^2
E. \eea

The quadratic action (\ref{eq:totalquadaction}) is invariant under the
foliation conserving transformation (\ref{eq:folitransf}),
%{\color{blue}(as like in)}
%{\color{blue} as it is in}
%Einstein gravity in which the theory is invariant under the full
%diffeomorphism,
so the action can be treated as a constraint system.
In a constraint system, it is important to classify  the constraints properly
and to count exact physical degrees of freedom. For this purpose,
Faddeev-Jackiw approach \cite{Faddeev:1988qp}, in which the
phase space can be reduced by solving the constraints and finally
the quadratic action can be expressed in terms only of the true physical
degrees of freedom,
seems to be powerful for linear analysis
%is more appropriate
and in this work we use
it to analyze the constraint system described by the quadratic
action (\ref{eq:totalquadaction}).

First, the conjugate momenta for the canonical variables $\psi, \delta \phi$ and $E$
are calculated by definition as
\bea%
\Pi^{\phi}&=&\frac{\delta(\delta_2S)}{\delta(\delta\phi^{\prime})}
=a^2(\delta\phi^{\prime}-\phi_0^{\prime}\Phi),\\
\Pi^\psi&=&\frac{\delta(\delta_2S)}{\delta\psi^{\prime}}
=a^2\frac{1-3\lambda}{\kappa^2}
[3\calh\Phi+\partial^2(\beta-E^{\prime})+3\psi^{\prime}]+3a^2\phi_0^\prime\delta\phi,
\label{eq:Pipsi_0}\\
\Pi^{E}&=&\frac{\delta(\delta_2S)}{\delta(\partial^2{E^{\prime}})}
=-a^2\frac{1-3\lambda}{\kappa^2}(\calh\Phi+\psi^{\prime})
-a^2\frac{1-\lambda}{\kappa^2}\partial^2(\beta-E^{\prime})-a^2\phi_0^{\prime}\delta\phi.
\eea%
Using these conjugate momenta, the quadratic action (\ref{eq:totalquadaction})
under the Legendre transformation becomes
\bea%
\delta_2 \cals=\int d^4x \Big[\Pi^\psi\psi^{\prime}+\Pi^{\phi}\delta
\phi^{\prime}
 + \Pi^{E} \partial^2 E^{\prime}
-\calh_c-\calc_0\Phi - \Phi \Sigma \Phi
-\calc_1\partial^2\beta\Big],
\label{eq:hamilquadact}
\eea%
where
\bea %
\calc_{0} &=& - \calh \Pi^{\psi} + \phi_0^{\prime} \Pi^{\phi} -
4 \{\xi a^2 \partial^2 - f_3 (\partial^2)^2 -  g_3 \frac{1}{a^2}
(\partial^2)^3 \} \psi + (a^4 V_{\phi} + 3 a^2 \calh
\phi_0^{\prime}) \delta \phi,
\\
\calc_{1} &=& \Pi^E \label{eq:constraint_c1},
\\
\calh_c &=&   \frac{(1-\lambda)\kappa^2}{4(1-3\lambda)a^2}
(\Pi^{\psi})^2
 + \frac{3\kappa^2}{4a^2} (\Pi^E)^2 + \frac{1}{2a^2} (\Pi^{\phi})^2
+ \frac{\kappa^2}{2a^2} \Pi^{\psi}\Pi^E
 - \frac{\kappa^2}{(1-3\lambda)}\phi_0^{\prime} \delta \phi \Pi^{\psi}
\nonumber \\
& & +  2\xi a^2 \psi \partial^2 \psi  +  (16f_1 +  6 f_2)(\partial^2
\psi)^2 - \frac{1}{a^2}(16 g_1 +
6g_2)\partial^2\psi(\partial^2)^2\psi
\nonumber \\
& & + a^4 \delta Z + \frac{1}{2}a^4 V_{\phi\phi} (\delta \phi)^2 +
\frac{3\kappa^2}{2(1-3\lambda)}a^2 \phi_0^{\prime 2} \delta \phi^2,
\\
\Sigma &=& a^2 \alpha \partial^2 - f_4 (\partial^2)^2 -
\frac{g_4}{a^2} (\partial^2)^3.
\eea%
As seen from (\ref{eq:hamilquadact}), $\partial^2 \beta$ appear linearly
without any time derivative, so its coefficient $\calc_1$ turns out the
primary constraint which is expected to vanish.
 $\calc_0$ is, however, not a constraint \cite{Gong:2010xp}, even if
  $\Phi$ does not have any time derivative terms in the action,
  because of the $\Phi$ squared term, $\Phi \Sigma \Phi$. Hence it should be determined from the
equation of motion for the auxiliary field $\Phi$,
\bea %
\frac{\delta(\delta_2 \cals)}{\delta \Phi(t^{\prime},y)} &=& \int d^4x
[-\calc_0 \delta(x-y)\delta(t-t^{\prime}) -2 \Sigma \Phi
\delta(x-y)\delta(t-t^{\prime})]
\nonumber \\
&=& -\calc_0 -2\Sigma \Phi = 0.
\eea%
Therefore we obtain % the $\Phi$ is
\bea%
\Phi = -\frac{\calc_0}{2\Sigma}. \label{eq:phisol}
\eea%

Following the Faddeev-Jackiw procedure\cite{Faddeev:1988qp},
we can reduce the phase space by solving the constraints ($\calc_1 =
\Pi^E = 0$),
then finally we obtain the quadratic action
including only physical degrees of freedom
\bea %
\delta_2 \cals = \int d^4x \biggl[\Pi^{\psi} \psi^{\prime} +
\Pi^{\phi} \delta \phi^{\prime} - \calh_{\ast} +
\frac{\calc_0^2}{4\Sigma} \biggr],\label{eq:lagrangian_red}
\eea%
where
\bea%
\calh_{\ast} &=&   \frac{(1-\lambda)\kappa^2}{4(1-3\lambda)a^2}
(\Pi^{\psi})^2  + \frac{1}{2a^2} (\Pi^{\phi})^2
 - \frac{\kappa^2}{(1-3\lambda)}\phi_0^{\prime} \delta \phi \Pi^{\psi}
\nonumber \\
& & +  2\xi a^2 \psi \partial^2 \psi  +  (16f_1 +  6 f_2)(\partial^2
\psi)^2 - \frac{1}{a^2}(16 g_1 +
6g_2)\partial^2\psi(\partial^2)^2\psi
\nonumber \\
& & + a^4 \delta Z + \frac{1}{2}a^4 V_{\phi\phi} (\delta \phi)^2 +
\frac{3\kappa^2}{2(1-3\lambda)}a^2 \phi_0^{\prime 2} \delta \phi^2.
\eea%
There is no more constraint so
we can not reduce the phase space further. As a result, we find that there are
two physical degrees of freedom.
This result is similar to that of the
projectable version of original Ho\v{r}ava-Lifshitz gravity
\cite{Gong:2010xp}, in which the lapse function $N$
 is a function of only time and
$\calc_0$ is also not a primary constraint, so there are two
physical degrees of freedom.

In order to compare with the conventional Hamiltonian formalism,
we count the
physical degrees of freedom from the quadratic action
(\ref{eq:hamilquadact}). From the beginning there are six
canonical variables {\it i.e.} $\psi, \phi, E$ and their conjugate momenta
and one
primary first class constraint $\calc_1$, so we have two physical
degrees of freedom ($ = \frac{1}{2}(6- 2\times 1)$),
that is $\psi$ and $\delta \phi$.

Next, we discuss the gauge invariant
quantities under the coordinate transformations, $x^{\mu}
\rightarrow x^{\mu} + \xi^{\mu}$. In (extended) Ho\v{r}ava-Lifshitz
gravity, $\xi^0 (t)$ is a function of only time, so the foliation
conserving diffeomorphism invariant symmetry is required instead of
the full
 diffeomorphism invariant symmetry under the gauge transformations.
 $\xi^i$ can be decomposed as a longitudinal component ($\partial^i \xi$)
 and a transverse component ($\xi_T^{i}$) in which $\xi_T^i$ satisfies
 $\partial_i \xi_T^i = 0$.
In the Hamiltonian constraint system, the first class constraints
play the role of the generators of the gauge
transformations\cite{Henneaux:1992ig}. In the present work,
because we have only one primary constraint, $\calc_1$ is
the only generator of the gauge transformations. 
Hence given perturbed quantity $f$, we can calculate the gauge transformation
as
 \bea \delta_{\xi}f = \{f,
\xi \calc_1 \}_{P}{\color{blue}},
\eea%
where the subscript $P$ denotes the Poisson brackets.
%we can calculate the gauge transformation of the canonical variables.
We find that
$\psi$ and $\delta \phi$ are the gauge invariant quantities
themselves ($\delta_{\xi} \psi = \delta_{\xi} \delta \phi = 0$)
 and $E$ is a pure gauge mode ($\delta_{\xi} E = \xi$).

The Hamilton's equations are
\bea%
\delta\phi^\prime &=& \{\delta \phi, \calh_T \}_P =
\frac{1}{a^2}\Pi^{\phi} - \frac{\phi_0^{\prime}}{2\Sigma}\calc_0,
\label{eq:eomphi}\\
\psi^\prime &=&
\frac{(1-\lambda)\kappa^2}{2(1-3\lambda)a^2}\Pi^{\psi}
-\frac{\kappa^2}{(1-3\lambda)}\phi_0^{\prime} \delta \phi +
\frac{\calh}{2\Sigma} \calc_0, \label{eq:eompsi}
\\
\Pi^{\phi\prime}
&=& \frac{\kappa^2}{(1-3\lambda)}\phi_0^{\prime} \Pi^{\psi}-a^4
\delta Z_{\delta \phi} -a^4 V_{\phi\phi} \delta \phi
-\frac{3\kappa^2}{(1-3\lambda)}a^2 \phi_0^{\prime 2} \delta \phi\nonumber\\
&& +\frac{1}{2\Sigma}(a^4 V_{\phi} + 3a^2 \calh
\phi_0^{\prime})\calc_0,\label{eq:eomPiphi}
\\
\Pi^{\psi\prime}
&=& -4\xi a^2 \partial^2 \psi - 4(8f_1+3f_2) (\partial^2)^2 \psi
+ \frac{4}{a^2}(8g_1 + 3g_2) (\partial^2)^3 \psi
\nonumber \\
& & +\frac{1}{2\Sigma} (-4\xi a^2 \partial^2 + 4f_3 (\partial^2)^2 +
\frac{4g_3}{a^2} (\partial^2)^3) \calc_0,\label{eq:eomPipsi}
\eea%
where
\bea
\calh_T = \calh_{\ast} - \calc_0^2/4\Sigma.
\eea
The second order differential equations for $\delta
\phi$ and $\psi$ are obtained from the Hamilton's equations
(\ref{eq:eomphi})-(\ref{eq:eomPipsi})
\bea%
& & \delta \phi^{\dprime} + 2\calh \delta \phi^{\prime}
+ \frac{\kappa^2}{(1-\lambda)} \phi_0^{\prime 2} \delta \phi
+ a^2 \delta Z_{\delta \phi} + a^2 V_{\phi\phi} \delta \phi
\nonumber \\
&=& \frac{2}{(1-\lambda)} \phi_0^{\prime}\psi^{\prime} - \frac{(1 -
3\lambda)}{(1-\lambda)} \calh\phi_0^{\prime} \Phi -2 a^2
V_{\phi}\Phi + \phi_0^{\prime} \Phi^{\prime},\\
& & \psi^{\dprime} + 2\calh \psi^{\prime}
+ \frac{2(1-\lambda)\kappa^2}{(1-3\lambda)a^2}
\biggl\{
 \xi a^2 \partial^2  + (8f_1+3f_2) (\partial^2)^2
- \frac{1}{a^2}(8g_1 + 3g_2) (\partial^2)^3 \biggr\} \psi
\nonumber \\
&=& -\frac{\kappa^2}{(1-3\lambda)}\phi_0^{\prime} \delta
\phi^{\prime} -\frac{\kappa^2}{(1-3\lambda)} (\phi_0^{\dprime} +
2\calh \phi_0^{\prime})
 \delta \phi
\nonumber \\
& &  -\frac{(1-\lambda)\kappa^2}{2(1-3\lambda)a^2}
\biggl\{ -4\xi a^2 \partial^2 + 4f_3 (\partial^2)^2
+ \frac{4g_3}{a^2} (\partial^2)^3 \biggr\} \Phi
\nonumber \\
& & -\calh \Phi^{\prime} - (2\calh^2 + \calh^{\prime}) \Phi.
\eea%
Here,  from (\ref{eq:phisol})  $\Phi$ becomes
\bea%
\Phi &=& \frac{\cald_3}{\cald_1}  \psi^{\prime}
 - \frac{a^2 \phi_0^{\prime}}{\cald_1} \delta \phi^{\prime}
- \frac{\cald_4}{\cald_1} \delta \phi + \frac{\cala}{\cald_1} \psi,
\eea%
with the functions defined by
\bea%
\cald_1 &=&  2\Sigma - \cald_2,
\\
\cald_2 &=&  \frac{2(1-3\lambda)a^2}{(1-\lambda)\kappa^2}\calh^2
+ a^2 \phi_0^{\prime 2},
\\
\cald_3 &=& \frac{2(1-3\lambda)a^2}{(1-\lambda)\kappa^2}\calh,
\\
\cald_4 &=&  \frac{(1-3\lambda)}{(1-\lambda)} a^2 \calh
\phi_0^{\prime} + a^4 V_{\phi},
\\
\cala &=& 4\{ \xi a^2 \partial^2 - f_3 (\partial^2)^2 - \frac{g_3}{a^2}
(\partial^2)^3 \}.
\eea%

Finally, we briefly discuss about the strong coupling of the scalar graviton
mode in the IR limit.
In the IR limit where the spatial derivative terms can be neglected,
$\Sigma$ and $\cala$ are negligible, so
 we get a relation $\cald_1 \simeq -\cald_2$.
 %The term $\cala$, therefore is negligible.
 The Lagrangian density of the action (\ref{eq:lagrangian_red})  then becomes
\bea %
\call &=& \biggl( \frac{(1-3\lambda)a^2}{(1-\lambda)\kappa^2} -
\frac{1}{2} \frac{\cald_3^2}{\cald_2} \biggr) \psi^{\prime 2}
+\frac{1}{2} a^2 \biggl( 1  - \frac{a^2 \phi_0^{\prime 2}}{\cald_2}
\biggr) \delta \phi^{\prime 2} + a^2 \phi_0^{\prime}
\frac{\cald_3}{\cald_2}\psi^{\prime} \delta \phi^{\prime}
\nonumber \\
& & + \biggl( \frac{\cald_3 \cald_4}{\cald_2}
 +  \frac{2a^2}{(1-\lambda)}\phi_0^{\prime}  \biggr)
\delta \phi \psi^{\prime}
 + \frac{1}{2} \biggl(a^2 \phi_0^{\prime} \frac{\cald_4}{\cald_2}
\biggr)^{\prime}  \delta \phi^2
 -\frac{1}{2} \frac{\cald_4^2}{\cald_2} \delta \phi^2
\nonumber \\
& &  -  2\xi a^2 \psi \partial^2 \psi  -  (16f_1 +  6 f_2)(\partial^2 \psi)^2
+ \frac{1}{a^2}(16 g_1 + 6g_2) \psi(\partial^2)^3\psi
\nonumber \\
& & - a^4 \delta Z - \frac{1}{2}a^4 V_{\phi\phi} (\delta \phi)^2
- \frac{\kappa^2 a^2}{2(1-\lambda)} \phi_0^{\prime 2} \delta \phi^2.
\label{eq:irlagrangian}
\eea%
In a general flat cosmological background, the scalar graviton mode $\psi$
is found to be coupled to the matter such as $\psi' \delta \phi'$ and
$\delta \phi \psi'$.
%In the limit $\lambda \rightarrow 1$, the dominant terms of $\cald_2$
%and $\cald_4$ are
%\bea%
%\cald_2 \simeq \frac{2(1-3\lambda)a^2}{(1-\lambda)\kappa^2}\calh^2,
%\quad \cald_4 \simeq \frac{(1-3\lambda)a^2}{(1-\lambda)}\calh \phi_0^{\prime},
%\eea%
%so the Lagrangian density contains the non-vanishing coupling terms
%%
%\bea%
%\call \supset \frac{a^2 \phi_0^{\prime}}{\calh}
%\psi^{\prime}  \delta \phi^{\prime}
%+ 3 a^2 \phi_0^{\prime} \delta \phi \psi^{\prime}.
%\eea%
%%
In addition, in the limit of $\lambda \rightarrow 1$
the coefficient of $\psi'^2$ is finite, so the scalar graviton mode
turns out to be propagating mode.

%This implies that even in the $\lambda \rightarrow 1$ limit, $\psi$
%can not decouple from the matter field, but is non-propagating mode
%because the coefficients of $\psi^{\prime 2}$ term cancel each other.

As a special case, we consider an inflationary accelerating background.
By taking the slow-roll approximations
%\bea%
($\dot{\phi}_0^{2} \ll H^2
\sim \kappa^2 V,~
\ddot{\phi_0} \ll 3H \dot{\phi}_0$ where $H = \frac{\dot{a}}{a}$ and dot
is a derivative with respect to  $t$),
%\eea%
 $\cald_2$ and $\cald_4$ approximate to
\bea %
\cald_2 \simeq
\frac{2(1-3\lambda)a^2}{(1-\lambda)\kappa^2}\calh^2,\quad
\cald_4 \simeq - \frac{2a^2}{(1-\lambda)}\calh \phi_0^{\prime}.
\eea%
The Lagrangian density (\ref{eq:irlagrangian}) then yields
\bea%
\call &=&  \frac{1}{2}a^2 \biggl(1-\frac{(1-\lambda)\kappa^2}{2(1-3\lambda)}
\frac{\phi_0^{\prime 2}}{\calh^2} \biggr) \delta \phi^{\prime 2}
%+ \frac{a^2 \phi_0^{\prime}}{\calh} \psi^{\prime} \delta \phi^{\prime}
- \frac{\kappa^2}{2(1-3\lambda)}\biggl(\frac{a^2 \phi_0^{\prime 2}}{\calh}
\biggr)^{\prime} \delta \phi^2
\nonumber \\
& &
 -  2\xi a^2 \psi \partial^2 \psi  -  (16f_1 +  6 f_2)(\partial^2 \psi)^2
+ \frac{1}{a^2}(16 g_1 + 6g_2) \psi(\partial^2)^3\psi
\nonumber \\
& & - a^4 \delta Z - \frac{1}{2}a^4 V_{\phi\phi} (\delta \phi)^2
-\frac{3\kappa^2 a^2}{2(1-3\lambda)}\phi_0^{\prime 2}\delta \phi^2.
\eea%
We ignore the coefficient of the term $\psi^{\prime} \delta \phi^{\prime}$,
 because
\bea%
\frac{a\phi_0^{\prime}}{\calh}
%= \frac{a\calh\phi_0^{\prime}}{\calh^2}
\propto \frac{m_p^2 V_{\phi}}{V} \ll 1.
\eea%
%
%in this approximation and then can be neglected.
Thus, the scalar graviton mode $\psi$ decouples from the matter
field completely and is also non-propagating mode because no kinetic
term of $\psi$ exists.

Although we discuss about the
strong coupling in the IR limit when $\lambda \rightarrow 1$ at the linear
order, in fact it seems to be necessary to go beyond linear order to tell
whether a mode of interest is strongly coupled or not
\cite{Mukohyama:2010xz}, which is out of the scope of the present paper.

%%%%%%%%%%%%%%%%%%%%%%%%%%%%%%%%%%%%%%%%%%%%%%%%%%%%%%%%%%%%%%%%%%%%
%\section{Conclusion}
%%%%%%%%%%%%%%%%%%%%%%%%%%%%%%%%%%%%%%%%%%%%%%%%%%%%%%%%%%%%%%%%%%%%

In this paper, we have investigated
the Hamiltonian structure of the extended version
of Ho\v{r}ava-Lifshitz gravity in a flat cosmological background
using Faddeev-Jackiw approach at linear order.
The gravitational potential in extension of
Ho\v{r}ava-Lifshitz gravity depends on
3-dimensional vector field $a_i$ constructed
from the lapse function $N(t,{\bf x})$
 as well as on 3-spatial metric $\gamma_{ij}$ and its derivative.

Since the quadratic action includes $\Phi^2$ term, \, $\calc_0$, the
coefficient of $\Phi$, is not a constraint, even if $\Phi$ does
not have any time derivative terms in the action.
 This implies that $\Phi$ could be determined from the equation
of motion. As a result,  there is only one primary constraint, $\calc_1$,
so we have two physical degrees of freedom and  one of them corresponds
to the scalar graviton mode.
This Hamiltonian structure is similar to that  of
the projectable version of original
Ho\v{r}ava-Lifshitz gravity \cite{Gong:2010xp} in which
the quadratic action includes $\Phi^2$ terms and $\calc_0 = 0$ in a
flat background so there are two physical degrees of freedom.

Further, we have found that in a general flat cosmological background
 the scalar
graviton mode is coupled to the scalar field and is propagating mode
 in the $\lambda \rightarrow 1$ limit. But as a special case,
 in an inflationary background with
the slow-roll approximation, it can be decoupled completely and
corresponds to the non-propagating mode.
But the linear order perturbations are not enough to talk about
the strong coupling issues, so
it would be interesting to go beyond linear order
in order to tell whether the mode is strongly coupled or not.
And although the linear instability issues
(see Ref. \cite{Mukohyama:2010xz} for a strong coupling
and a linear instability in original Ho\v{r}ava-Lifshitz gravity
for a cosmological background) are not treated in this paper,
it would also be important questions to be answered in
the extended version of Ho\v{r}ava-Lifshitz gravity.

%%%%%%%%%%%%%%%%%%%%%%%%%%%%%%%%%%%%%%%%%%%%%%%%%%%%%%%%%%%%%%%%%%%%
\acknowledgments
We thank Mu-In Park and Frederico Arroja for useful comments and discussions.
We also thank Shinji Mukohyama for useful comments, especially for
pointing out the strong coupling problem in the IR limit at the nonlinear
order.
SK thanks
the Yukawa Institute  for Theoretical Physics at Kyoto University,
where this work was completed during the YITP-T-10-01 on "Gravity and
Cosmology (GC2010)".
SK is supported by the National Research Foundation of Korea Grant funded
by the Korean Government [NRF-2009-353-C00007]. This work was supported by
the National Research Foundation of Korea(NRF) grant funded by the Korea
government(MEST) through the Center for Quantum Spacetime(CQUeST) of Sogang
University with grant number 2005-0049409 (S. S.).

%%%%====================bibliography=================================


\begin{thebibliography}{99}

%\cite{Horava:2009uw}
\bibitem{Horava:2009uw}
  P.~Horava,
  %``Quantum Gravity at a Lifshitz Point,''
  Phys.\ Rev.\  D {\bf 79}, 084008 (2009)
  [arXiv:0901.3775 [hep-th]].
  %%CITATION = PHRVA,D79,084008;%%

%\cite{Lu:2009em}
\bibitem{Lu:2009em}
  H.~Lu, J.~Mei and C.~N.~Pope,
  %``Solutions to Horava Gravity,''
  Phys.\ Rev.\ Lett.\  {\bf 103}, 091301 (2009)
  [arXiv:0904.1595 [hep-th]];
  %%CITATION = PRLTA,103,091301;%%

%\cite{Cai:2009pe}
%\bibitem{Cai:2009pe}
  R.~G.~Cai, L.~M.~Cao and N.~Ohta,
  %``Topological Black Holes in Horava-Lifshitz Gravity,''
  Phys.\ Rev.\  D {\bf 80}, 024003 (2009)
  [arXiv:0904.3670 [hep-th]];
  %%CITATION = PHRVA,D80,024003;%%

%\cite{Kehagias:2009is}
%\bibitem{Kehagias:2009is}
  A.~Kehagias and K.~Sfetsos,
  %``The black hole and FRW geometries of non-relativistic gravity,''
  Phys.\ Lett.\  B {\bf 678}, 123 (2009)
  [arXiv:0905.0477 [hep-th]].
  %%CITATION = PHLTA,B678,123;%%

  %\cite{Calcagni:2009ar}
\bibitem{Calcagni:2009ar}
  G.~Calcagni,
  %``Cosmology of the Lifshitz universe,''
  JHEP {\bf 0909}, 112 (2009)
  [arXiv:0904.0829 [hep-th]];
  %%CITATION = JHEPA,0909,112;%%

%\cite{Kiritsis:2009sh}
%\bibitem{Kiritsis:2009sh}
  E.~Kiritsis and G.~Kofinas,
  %``Horava-Lifshitz Cosmology,''
  Nucl.\ Phys.\  B {\bf 821}, 467 (2009)
  [arXiv:0904.1334 [hep-th]];
  %%CITATION = NUPHA,B821,467;%%

%\cite{Mukohyama:2009gg}
%\bibitem{Mukohyama:2009gg}
  S.~Mukohyama,
  %``Scale-invariant cosmological perturbations from Horava-Lifshitz gravity
  %without inflation,''
  JCAP {\bf 0906}, 001 (2009)
  [arXiv:0904.2190 [hep-th]].
  %%CITATION = JCAPA,0906,001;%%

%\cite{Mukohyama:2010xz}
\bibitem{Mukohyama:2010xz}
  S.~Mukohyama,
  %``Horava-Lifshitz Cosmology: A Review,''
  arXiv:1007.5199 [hep-th].
  %%CITATION = ARXIV:1007.5199;%%

  %\cite{Park:2009zr}
\bibitem{Park:2009zr}
  M.~i.~Park,
  %``A Test of Horava Gravity: The Dark Energy,''
  JCAP {\bf 1001}, 001 (2010)
  [arXiv:0906.4275 [hep-th]];
  %%CITATION = JCAPA,1001,001;%%

%\cite{Dutta:2009jn}
%\bibitem{Dutta:2009jn}
  S.~Dutta and E.~N.~Saridakis,
  %``Observational constraints on Horava-Lifshitz cosmology,''
  JCAP {\bf 1001}, 013 (2010)
  [arXiv:0911.1435 [hep-th]];
  %%CITATION = JCAPA,1001,013;%%

%\cite{Kim:2009dq}
%\bibitem{Kim:2009dq}
  S.~S.~N.~Kim, T.~Kim and Y.~Kim,
  %``Surplus Solid Angle: Toward Astrophysical Test of Horava-Lifshitz
  %Gravity,''
  Phys.\ Rev.\  D {\bf 80}, 124002 (2009)
  [arXiv:0907.3093 [hep-th]];
  %%CITATION = PHRVA,D80,124002;%%

%\cite{Mukohyama:2009zs}
%\bibitem{Mukohyama:2009zs}
  S.~Mukohyama, K.~Nakayama, F.~Takahashi and S.~Yokoyama,
  %``Phenomenological Aspects of Horava-Lifshitz Cosmology,''
  Phys.\ Lett.\  B {\bf 679}, 6 (2009)
  [arXiv:0905.0055 [hep-th]];
  %%CITATION = PHLTA,B679,6;%%

%\cite{Takahashi:2009wc}
%\bibitem{Takahashi:2009wc}
  T.~Takahashi and J.~Soda,
  %``Chiral Primordial Gravitational Waves from a Lifshitz Point,''
  Phys.\ Rev.\ Lett.\  {\bf 102}, 231301 (2009)
  [arXiv:0904.0554 [hep-th]];
  %%CITATION = PRLTA,102,231301;%%

%\cite{Koh:2009cy}
%\bibitem{Koh:2009cy}
  S.~Koh,
  %``Relic gravitational wave spectrum, the trans-Planckian physics and
  %Ho\v{r}ava-Lifshitz gravity,''
  arXiv:0907.0850 [hep-th].
  %%CITATION = ARXIV:0907.0850;%%

%\cite{Blas:2009yd}
\bibitem{Blas:2009yd}
  D.~Blas, O.~Pujolas and S.~Sibiryakov,
  %``On the Extra Mode and Inconsistency of Horava Gravity,''
  JHEP {\bf 0910}, 029 (2009)
  [arXiv:0906.3046 [hep-th]].
  %%CITATION = JHEPA,0910,029;%%


%\cite{Li:2009bg}
\bibitem{Li:2009bg}
  M.~Li and Y.~Pang,
  %``A Trouble with Ho\v{r}ava-Lifshitz Gravity,''
  JHEP {\bf 0908}, 015 (2009)
  [arXiv:0905.2751 [hep-th]];
  %%CITATION = JHEPA,0908,015;%%



%\cite{Koyama:2009hc}
%\bibitem{Koyama:2009hc}
  K.~Koyama and F.~Arroja,
  %``Pathological behaviour of the scalar graviton in Ho\v{r}ava-Lifshitz
  %gravity,''
  arXiv:0910.1998 [hep-th];
  %%CITATION = ARXIV:0910.1998;%%

%\cite{Henneaux:2009zb}
%\bibitem{Henneaux:2009zb}
  M.~Henneaux, A.~Kleinschmidt and G.~L.~Gomez,
  %``A dynamical inconsistency of Horava gravity,''
  arXiv:0912.0399 [hep-th].
  %%CITATION = ARXIV:0912.0399;%%


%\cite{Papazoglou:2009fj}
%\bibitem{Papazoglou:2009fj}
  A.~Papazoglou and T.~P.~Sotiriou,
  %``Strong coupling in extended Horava-Lifshitz gravity,''
  Phys.\ Lett.\  B {\bf 685}, 197 (2010)
  [arXiv:0911.1299 [hep-th]];
  %%CITATION = PHLTA,B685,197;%%

%\cite{Blas:2009ck}
%\bibitem{Blas:2009ck}
  D.~Blas, O.~Pujolas and S.~Sibiryakov,
  %``Comment on `Strong coupling in extended Horava-Lifshitz gravity',''
  arXiv:0912.0550 [hep-th].
  %%CITATION = ARXIV:0912.0550;%%

%\cite{Blas:2009qj}
\bibitem{Blas:2009qj}
  D.~Blas, O.~Pujolas and S.~Sibiryakov,
  %``Consistent Extension Of Horava Gravity,''
  Phys.\ Rev.\ Lett.\  {\bf 104}, 181302 (2010)
  [arXiv:0909.3525 [hep-th]].
  %%CITATION = PRLTA,104,181302;%%


  %\cite{Kluson:2010nf}
\bibitem{Kluson:2010nf}
  J.~Kluson,
  %``Note About Hamiltonian Formalism of Healthy Extended Horava-Lifshitz
  %Gravity,''
  JHEP {\bf 1007}, 038 (2010)
  [arXiv:1004.3428 [hep-th]];
  %%CITATION = JHEPA,1007,038;%%

%\cite{Kimpton:2010xi}
%\bibitem{Kimpton:2010xi}
  I.~Kimpton and A.~Padilla,
  %``Lessons from the decoupling limit of Horava gravity,''
  JHEP {\bf 1007}, 014 (2010)
  [arXiv:1003.5666 [hep-th]];
  %%CITATION = JHEPA,1007,014;%%

%\cite{Blas:2009ck}
%\bibitem{Blas:2009ck}
  D.~Blas, O.~Pujolas and S.~Sibiryakov,
  %``Comment on `Strong coupling in extended Horava-Lifshitz gravity',''
  Phys.\ Lett.\  B {\bf 688}, 350 (2010)
  [arXiv:0912.0550 [hep-th]];
  %%CITATION = PHLTA,B688,350;%%

%\cite{Papazoglou:2009fj}
%\bibitem{Papazoglou:2009fj}
  A.~Papazoglou and T.~P.~Sotiriou,
  %``Strong coupling in extended Horava-Lifshitz gravity,''
  Phys.\ Lett.\  B {\bf 685}, 197 (2010)
  [arXiv:0911.1299 [hep-th]].
  %%CITATION = PHLTA,B685,197;%%


%\cite{Blas:2010hb}
\bibitem{Blas:2010hb}
  D.~Blas, O.~Pujolas and S.~Sibiryakov,
  %``Models of non-relativistic quantum gravity: the good, the bad and the
  %healthy,''
  arXiv:1007.3503 [hep-th].
  %%CITATION = ARXIV:1007.3503;%%


%\cite{Gong:2010xp}
\bibitem{Gong:2010xp}
  J.~O.~Gong, S.~Koh and M.~Sasaki,
  %``A complete analysis of linear cosmological perturbations in
  %Ho\v{r}ava-Lifshitz gravity,''
  arXiv:1002.1429 [hep-th].
  %%CITATION = ARXIV:1002.1429;%%


%\cite{Faddeev:1988qp}
\bibitem{Faddeev:1988qp}
  L.~D.~Faddeev and R.~Jackiw,
  %``Hamiltonian Reduction of Unconstrained and Constrained Systems,''
  Phys.\ Rev.\ Lett.\  {\bf 60}, 1692 (1988).
  %%CITATION = PRLTA,60,1692;%%


%\cite{Kobayashi:2010eh}
\bibitem{Kobayashi:2010eh}
  T.~Kobayashi, Y.~Urakawa and M.~Yamaguchi,
  %``Cosmological perturbations in a healthy extension of Horava gravity,''
  JCAP {\bf 1004}, 025 (2010)
  [arXiv:1002.3101 [hep-th]].
  %%CITATION = JCAPA,1004,025;%%

%\cite{Henneaux:1992ig}
\bibitem{Henneaux:1992ig}
  M.~Henneaux and C.~Teitelboim,
  ``Quantization of gauge systems,''
%\href{http://www.slac.stanford.edu/spires/find/hep/www?irn=2824396}{SPIRES entry}
{\it  Princeton, USA: Univ. Pr. (1992) 520 p}


%%\cite{Anderegg:1994xq}
%\bibitem{Anderegg:1994xq}
%  S.~Anderegg and V.~F.~Mukhanov,
%  %``Path integral quantization of cosmological perturbations,''
%  Phys.\ Lett.\  B {\bf 331}, 30 (1994)
%  [arXiv:hep-th/9403091];
%  %%CITATION = PHLTA,B331,30;%%

%%\cite{Garriga:1997wz}
%%\bibitem{Garriga:1997wz}
%  J.~Garriga, X.~Montes, M.~Sasaki and T.~Tanaka,
%  %``Canonical quantization of cosmological perturbations in the one-bubble
%  %open universe,''
%  Nucl.\ Phys.\  B {\bf 513}, 343 (1998)
%  [arXiv:astro-ph/9706229].
 % %%CITATION = NUPHA,B513,343;%%












\end{thebibliography}
\end{document}